\documentclass[aps, prx,twocolumn,superscriptaddress,showpacs,amsmath,amssymb]{revtex4-1}

\usepackage{amssymb}
\usepackage[dvips]{epsfig,graphics,color}
\usepackage{multirow}
\usepackage{hhline}
\usepackage{graphicx}

\usepackage{color}
\usepackage[normalem]{ulem}

\usepackage{color}

\newcommand{\COMMENTED}[1]{}

\emergencystretch=\maxdimen
\begin{document}

\title{Numerical results on the short-range spin correlation functions \\
in the ground state of the two-dimensional Hubbard model}

\author{Mingpu Qin}
\affiliation{Department of Physics, College of William and Mary, Williamsburg, Virginia 23187}

\author{Hao Shi}
\affiliation{Department of Physics, College of William and Mary, Williamsburg, Virginia 23187}

\author{Shiwei Zhang}
\affiliation{Department of Physics, College of William and Mary, Williamsburg, Virginia 23187}

\begin{abstract}
Optical lattice experiments with ultracold fermion atoms and quantum gas microscopy 
have recently realized direct measurements of magnetic correlations at the site-resolved level.
We calculate the short-range spin correlation functions in the ground state of the two-dimensional repulsive Hubbard model with the auxiliary-field Quantum Monte Carlo (AFQMC) method. The results are numerically exact at half filling where 
the fermion sign problem is absent. Away from half-filling, we employ
the constrained path AFQMC approach to eliminate the exponential computational scaling from the sign problem. 
The constraint  employs unrestricted Hartree-Fock trial wave-functions
with an effective interaction strength $U$, which is optimized self-consistently within AFQMC.
Large supercells are studied, with twist averaged boundary conditions as needed, to reach 
the thermodynamic limit. We find that
 the nearest-neighbor spin correlation always increases with the interaction strength $U$, contrary to the finite-temperature behavior where a maximum is reached at a finite 
$U$ value. We also observe a change of sign in the next nearest neighbor spin correlation with increasing 
density,
which
is a consequence of the buildup of the long-range anti-ferromagnetic correlation. We expect the results presented in this work to serve as a 
benchmark  as lower temperatures are reached in ultracold atom experiments. 
\end{abstract}


\maketitle

\section{introduction}
The Hubbard model \cite{Hubbard_origional} is one of the most studied models in physics.
The model is believed to be relevant to many correlated electron phenomena including interaction-driven metal-insulator
transitions \cite{Imada_rmp_1998}, magnetism \cite{ph-sym} and spin and charge density waves \cite{cdw_sdw}, and most importantly
high-temperature superconductivity \cite{htc}. Except in one dimension \cite{lieb_wu}, however, there is no analytic solution to the
model. Numerical simulations have thus played an increasingly larger role in the study of the Hubbard model \cite{paper_simons}.

The development in ultracold atom experiments provides another possibility 
for direct ``simulation'' of the Hubbard model  \cite{rev_zoller,rev_bloch}.
Recently, the two-dimensional Hubbard model was realized with ultracold atoms in optical lattices,
and a flurry of activities have been reported
 where both local quantities and short-range (spin and charge) correlations were measured
\cite{nat_519_2015,prl_116_175301,prl_116_235301,sci_1253,sci_1257,sci_1260}.
Numerical results have 
usually been used to benchmark the experimental results.
In addition to serving as
a thermometry for ultracold atoms, computational results have been integrated in these studies to 
guide the experiments and interpretation.

Understanding the properties of the doped Hubbard model is challenging,
because of
the existence of different competing orders for the ground state which are separated by tiny energy scales \cite{stripe}.
High accuracy and resolution is required to distinguish and characterize the different candidate states.
To date, the temperature that can be reached by ultracold atom experiments is still relatively ``high'' compared to the energy scales of the competing ground-state orders.
In this regime, reliable 
numerical results have been provided by a multitude of computational 
methods including finite temperature determinant quantum Monte Carlo (DQMC) \cite{prl_104_066406,prl_106_035301}, dynamical cluster approximation
(DCA) \cite{prb_88_155108}, and numerical linked-cluster expansion (NLCE) \cite{pra_84_053611,prl_109_205301}, all of which work at finite temperatures.
 
 As experiments continue the development of cooling technology, we can expect lower temperatures to be reached in the near future. In fact 
a lower temperature of $T = 0.25(2)t$ has already been reported in the most recent experiment \cite{Mazurenko_2016}.
 This offers the exciting prospect of determining and understanding the low-temperature and ground-state
phases of the Hubbard model by the optical lattice experiments.  
Reliable numerical
data at lower temperatures and ground state will be crucial for benchmarking experimental results and assisting interpretation and analysis. 
Reaching lower temperatures present significant challenges \cite{paper_simons}
for numerical methods, however, and numerical data will be less readily available or reliable.

In anticipation of these developments and challenges, we 
calculate the short-range spin correlation functions of the 2D Hubbard model at zero temperature. 
Applying the state-of-the-art auxiliary-field
quantum Monte Carlo (AFQMC) method, we study systems with large size and take advantage of twist averaged boundary conditions to reach
the thermodynamic limit. At half-filling, where the sign problem \cite{sign_1,sign_2} is absent, the results are numerically exact. 
For doped systems, we employ the constrained-path (CP) approximation  \cite{zhang_prb_1997} to
deal with the sign problem. In CP-AFQMC the sign problem is eliminated by a constraint on the random 
walks in Slater determinant space which is  dependent on the trial wave-function.
Previous studies have shown that this bias is usually small \cite{chia-chen_EOS,CPMC_sym_1}. 
A further recent advance  allows a self-consistent optimization of the trial wave-function  \cite{prb_self-consistent}.
For each filling and interaction strength,
we test a series of trial wave-functions generated from unrestricted Hartree-Fock (UHF)  with different $U$ values, relying on the feedback from the CP-AFQMC calculation to self-consistently determine the 
optimal effective $U$ value, $U_{\rm eff}^{\rm UHF}$, for the UHF.
As a result, the computational approach we use can treat sufficiently large system sizes in the ground 
state, and 
obtain highly accurate results \cite{paper_simons,stripe}

We find major differences in the behavior of the short-range magnetic correlations from what has been 
observed to date experimentally and computationally. At half-filling, 
 the spin
correlation always increases with the interaction strength $U$ and  monotonically  approaches the Heisenberg limit with $U\rightarrow \infty$. This behavior
of the spin correlation is in sharp contrast with the finite-temperature situation, where the correlation
reaches its maximum at a finite $U$. For the more important doped cases, we observe a change of sign in the next nearest
neighboring (NNN) correlation functions which is a precursor for the onset of 
spin-density wave orders and the
 buildup of the anti-ferromagnetic correlation with increase of density. 
We find that the filling factor where the NNN correlations function changes
sign is essentially independent of interaction strengths.            

The rest of the paper is organized as follows. In Sec.~\ref{model_method}, we first define the Hubbard model
and give a brief summary of the method used in this work. 
We will comment on the approach to the thermodynamic limit with twist averaged boundary conditions (TABC), and mention some of the latest 
technical advances employed in the study to improve accuracy and computational capabilities and efficiency. 
In Sec.~\ref{half_filling} we present the results on  spin correlations at half-filling as a function of $U$ values. 
In Sec.~\ref{away_half_filling}, the nearest and next nearest neighboring spin correlation functions for systems away from half-filling are presented. A short summary
in Sec.~\ref{summary} will conclude this paper.

\section{Model and Method}
\label{model_method}
\subsection{Hubbard Model}
The Hubbard model is defined as
\begin{equation}
	H = K + V = -\sum\limits_{ i,j, s}t_{ij}  \left(c_{i,s}^\dagger c_{j,s}  + H.c.\right) +U\sum\limits_i n_{i\uparrow}n_{i\downarrow},
	\label{eqn:H}
\end{equation}
where $K$ and $V$ are the kinetic and on-site interaction terms, respectively. 
The creation (annihilation) operator on site $i$ is
$c_{i,s}^\dagger$ ($c_{i,s}$), with $s = \uparrow,\downarrow$ the spin
of the electron,
and $n_{i,s}$ is the corresponding number operator.
We denote the total number of electrons with $ \uparrow$- and $\downarrow$-spin by
$N_\uparrow$ and $N_\downarrow$, respectively. In this work, only spin-balanced systems
($N_\uparrow = N_\downarrow$) are considered. The filling factor
is defined as $n = (N_{\uparrow}+N_{\downarrow} )/{N}$ with
$N$ being the total number of lattice sites in the supercell. So $n=1$ means half-filling.
We will typically use supercells of square lattice with size $N = L \times L$.
We only consider nearest neighbor and uniform hopping in this work, $t_{ij}=t$ for each 
near-neighbor pair $\langle ij\rangle$, and $t$ is set as the energy unit. 
The strength of the repulsive interaction is given by $U/t$.

The quantity we investigate in this work is the spin correlation function,
\begin{equation}
c(\mathbf{r})=\langle\psi_{g}|\overrightarrow{S}(\mathbf{0})\cdot \overrightarrow{S}(\mathbf{r})|\psi_{g}\rangle,
\label{eqn:s(r)}
\end{equation}
where $|\psi_g\rangle$ is the ground state of $H$ in Eq.~(\ref{eqn:H}).
The spin operator at  $\mathbf{r}$ is given by
\begin{equation}
\overrightarrow{S}(\mathbf{r})=\frac{1}{2}\sum_{ss^{\prime}}c_{i,s}^{\dagger}\overrightarrow{\sigma}_{s,s^{\prime}}c_{i,s^{\prime}},
\end{equation}
 where $i$ is the lattice site label of the position $\mathbf{r}$,
and $\overrightarrow{\sigma}$ are the Pauli matrices.
Because of translational invariance, the reference site `$\mathbf{0}$' can be averaged over in the periodic
supercell, and the correlation function is only a function of lattice vector ${\mathbf r}$ and satisfies all lattice symmetries.

In order to extrapolate more reliably to the  thermodynamic limit (TDL), we adopt
twist averaged boundary conditions \cite{TBC}.
Under twist boundary conditions (TBC), an electron
gains a phase when hopping across the boundaries:
\begin{equation}
\Psi(\ldots,\mathbf{r}_m+\mathbf{L},\ldots) = e^{i\hat{\mathbf{L}}\cdot\mathbf{\Theta}}\Psi(\ldots,\mathbf{r}_m,\ldots),
	\label{eqn:tbc}
\end{equation}
where $\hat{\mathbf{L}}$ is the unit vector along $\mathbf{L}$, 
$\mathbf{r}_m$ is the position of the $m$-th electron, 
and
the twist angle $\mathbf{\Theta}=(\theta_x,\theta_y)$ is a two dimensional
parameter, with $\theta_x$  ($\theta_y$)  $\in [0, 2\pi)$. 
This is equivalent to placing the lattice on a torus and applying a 
magnetic field which induces a flux of $\theta_x$ along the $x$-direction (and a flux of $\theta_y$ along the $y$-direction).
In Eq.~(\ref{eqn:tbc}), the 
translational symmetry is explicitly broken, however it can be easily restored with
an alternative gauge, i.e., adjust $t$ to
$t \times e^{i\theta_x / L}$   along $x$ ($t \times e^{i\theta_y / L}$ along $y$).

To implement TABC, a set of  $N_\theta$ twist angles is chosen and we carry out an independent calculation for each twist.
The final result of a physical quantity is the averaged value from these independent calculations. 
The constrained path condition can be generalized straightforwardly to the case of TBC \cite{chia-chen_EOS}.
By imposing a random TBC, the possible degeneracy of the non-interacting energy levels
is lifted by explicitly breaking the rotational symmetry of the lattice. This eliminates 
the so called open-shell effects and helps to reduce the constraint bias when a free-electron trial wave function
is used. In most of our calculations, we use UHF trial wave functions so that this is not 
especially relevant. Nevertheless we have retained the same procedure.

When implementing the TABC, we use a quasi-random instead of a pseudo-random
sequence to generate the twist. As shown in Ref.~\cite{prb_benchmark}, the TABC physical quantities  converge faster (with respect to 
the number of twists used) with quasi-random twists than with a pseudo-random sequence. A quasi-random sequence is  distributed more uniformly in the sampled space. 
Twists from a uniform grid have essentially the same convergence rate as a quasi-random sequence, however it 
can lead to open-shell situations in the many-body calculation, which is less advantageous especially when a 
simple trial wave function is used \cite{prb_benchmark}.  Additionally, a uniform grid is typically not cumulative which means that, 
 if the targeted statistical accuracy is not reached with a given set,
we would likely not be able to re-use the data, 
and new calculations would be necessary for each twist in the new set. A quasi-random sequence
avoids these problems and combines 
the advantages of a pseudo-random sequence and a uniform grid.

\subsection{Auxiliary-field quantum Monte Carlo method}
\label{ssec:method}

In this section, we will briefly introduce the methods used in this work.
(A more comprehensive
discussion of this method can be found in Ref.~\cite{lecture-notes}.)  
Our AFQMC calculations share much of the same basic framework as 
standard DQMC method \cite{AFQMC} and its ground-state variant \cite{koonin}. 
By repeatedly applying the projection operator to a state $|\psi_0\rangle$ with non-zero
overlap with the ground state $|\psi_g\rangle$ of the Hamiltonian $H$ in Eq.~(\ref{eqn:H}), we can obtain $|\psi_g\rangle$:
\begin{equation}
	|\psi_{g}\rangle \propto \lim_{\beta \rightarrow\infty}e^{-\beta H}|\psi_{0}\rangle
	\label{eqn:proj}
\end{equation}
and the expectation value of an operator $O$ can be represented as
\begin{equation}
	\langle O\rangle=\frac{\langle\psi_{0}|e^{-\beta H}Oe^{-\beta H}|\psi_{0}\rangle}{\langle\psi_{0}|e^{-2\beta H}|\psi_{0}\rangle}\,.
	\label{eqn:expect_O}
\end{equation}

Through the Trotter Suzuki decomposition, the kinetic and interaction parts in the projection operator can be decoupled as:
\begin{equation}
	e^{-\beta H}=(e^{-\tau H})^{n}=(e^{-\frac{1}{2}\tau K}e^{-\tau V}e^{-\frac{1}{2}\tau K})^{n}+O(\tau^{2}),
\end{equation}
where $\beta = \tau n$. The Trotter error can be eliminated by an extrapolation of $\tau$ to $0$. 
We typically choose $\tau = 0.01$ in this work. We have verified that the Trotter error with $\tau = 0.01$ is smaller than
the targeted statistical errors.

The initial state $|\psi_0\rangle$ is usually chosen as a Slater determinant in AFQMC. The one-body
term $e^{-\frac{1}{2}\tau K}$ can be directly applied to it and 
the result is another Slater determinant. This is not true for the two-body term $e^{-\tau V}$. However,
the two-body term can be decomposed into an integral of one-body terms through the so-called Hubbard-Stratonovich (HS) transformation. Different
types of HS transformations \cite{TREMBLAY_1992} for $e^{-\tau V}$ exist in literature. The two commonly used types are the so called spin decomposition  
\begin{equation}
	e^{-{\tau}Un_{\uparrow}n_{\downarrow}}=e^{-{\tau}U(n_{\uparrow}+n_{\downarrow})/2}\sum_{x=\pm1}\frac{1}{2}e^{\gamma_s x(n_{\uparrow}-n_{\downarrow})}\,,
	\label{eq:spin-decom}
\end{equation} 
with the constant $\gamma_s$ determined by $\cosh(\gamma_s) \equiv  \exp({\tau}U/2)$, and the charge decomposition
\begin{equation}
	e^{-{\tau}Un_{\uparrow}n_{\downarrow}}=e^{-{\tau}U(n_{\uparrow}+n_{\downarrow}-1)/2}\sum_{x=\pm1}\frac{1}{2}e^{\gamma_c x(n_{\uparrow}+n_{\downarrow}-1)}\,,
	\label{eq:charge-decomp}
\end{equation}
with
$\cosh(\gamma_c) \equiv \exp(-{\tau}U/2)$ \cite{hirsch_prb_1983}. Here $x$ is an Ising-spin-like auxiliary field. 
Transformations with continuous Gaussian auxiliary-fields exist which can be made  \cite{CPMC_sym_1} 
essentially as efficient
as the discrete forms.
Different choices of the HS can lead to different accuracies or efficiencies, because of symmetry 
considerations \cite{CPMC_sym_1,CPMC_sym_2} or other factors \cite{Hao-inf-var}.

With the HS transformation,
Eq.~(\ref{eqn:expect_O}) turns into
\begin{equation}
	\langle O\rangle=\frac{\sum_{\{X_{i},X_{j}\}}\langle\psi_{0}|\prod_{i=1}^{n}P_{i}(X_{i})O\prod_{j=1}^{n}P_{j}(X_{j})|\psi_{0}\rangle}{\sum_{\{X_{i},X_{j}\}}\langle\psi_{0}|\prod_{i=1}^{n}P_{i}(X_{i})\prod_{j=1}^{n}P_{j}(X_{j})|\psi_{0}\rangle}
	\label{eq:obs}
\end{equation}
where $X_{i}$ is the collection of the $N$ auxiliary fields from the HS transformation, and $P_i$ is the product of the kinetic term 
and the one-body terms from the HS transformation at time slice $i$.
The multi-dimensional integrals 
can then be evaluated by 
Monte Carlo methods, e.g., with the Metropolis algorithm.

At half-filling,
each individual term in the sum in the denominator of Eq.~(\ref{eq:obs}) is
always non-negative, because of  particle-hole symmetry \cite{ph-sym}. So we can use it as probability density and
the sign problem is absent. 
Our calculations in this work use mostly the 
 path integral form outlined above, but introduce several recent algorithmic advances such as 
 an acceleration technique \cite{hao_pra_2015} (with force bias \cite{lecture-notes,zhang_prl_2003} in the Metropolis sampling) and control of the divergence of Monte Carlo variance \cite{Hao-inf-var}. 

Away from half-filling, a direct evaluation of Eq.~(\ref{eq:obs}) by Monte Carlo will suffer from the  sign problem \cite{sign_1,sign_2},
since terms in 
the denominator of Eq.~(\ref{eq:obs}) can 
become negative for some auxiliary fields \cite{lecture-notes}.
The sign problem can be eliminated by the constrained path approximation. The framework within which 
this has been implemented in Hubbard-like model has often been referred to as the constrained path 
Monte Carlo (CPMC) method  \cite{zhang_prb_1997}. Here we will refer to the approach as CP-AFQMC
to be consistent with recent conventions.
A description of the 
CP-AFQMC method as applied to Hubbard-like models can be found in Ref.~\cite{Huy_CPC_2014}.

In CP-AFQMC, the wave function is represented as an ensemble of a set of Slater determinants 
which are 
called walkers. The evolution of wave function in the imaginary time is represented as random walks in the Slater determinant space by 
sampling the auxiliary fields. 
The paths of auxiliary-fields are constrained to ensure the
overlap of any propagated walker with the trial wave-function, $|\psi_T\rangle$, computed at each time slice, remains non-negative. 
Physical quantities can be evaluated using the mixed estimate as
\begin{equation}
	\langle O\rangle_{\rm mixed}=\frac{\sum_{k}w_{k}\langle\psi_{T}|O|\psi_{k}\rangle}{\sum_{k}w_{k}\langle\psi_{T}|\psi_{k}\rangle}\,,
	\label{mix_estimate}
\end{equation}
where 
$|\psi_{k}\rangle$ is the kth walker, and
$w_k$ is the corresponding weight.
The mixed estimate is used to compute the energy (and other observables which 
commute with the Hamiltonian). For observables which do not commute with the Hamiltonian, the mixed estimate
is biased, and 
we use back propagation to correct for this \cite{zhang_prb_1997, Wirawan-PRE}.

As mentioned, the CP-AFQMC  overcomes the sign problem and restores a low
computational scaling with system size, 
at the cost of a systematic error which depends on the
trial wave-function $|\psi_T\rangle$. 
Previous studies have shown the systematic error is small even with a free-electron or Hartree-Fork trial wave-function \cite{chia-chen_EOS}.
In this work, we adopt a recent advance \cite{prb_self-consistent} which allows the self-consistent 
construction of an optimal 
optimal UHF trial wave-function 
with an effective interaction $U_{\rm eff}^{\rm UHF}$.
In the self-consistent procedure, we couple the CP-AFQMC calculation
with a mean-field calculation. We first carry out an ordinary
CP-AFQMC calculation with a free electron or UHF trial wave-function. Then instead of solving the mean-field Hamiltonian self-consistently, we
feed the local density from the CP-AFQMC calculation as the ``field" in the mean-field Hamiltonian and scan the interaction $U$ to find the solution
which yields a local density closest to the input density. Next we use the mean field solution with this ``optimal'' $U$ as the trial wave-function for the next-step
CP-AFQMC calculation. The same process is repeated until the local density is converged.

 The use of the self-consistent CP-AFQMC further improved the accuracy, especially for determining 
spin- and charge-orders \cite{prb_self-consistent}. 
We have carried out additional benchmarks here in smaller systems specifically targeting short-range correlations. Therefore results presented here, consistent with previous experience \cite{paper_simons,stripe},
are expected to be very accurate.

\section{Results at half-filling}
\label{half_filling}

 \begin{figure}[t]
 	\includegraphics[width=8.06cm]{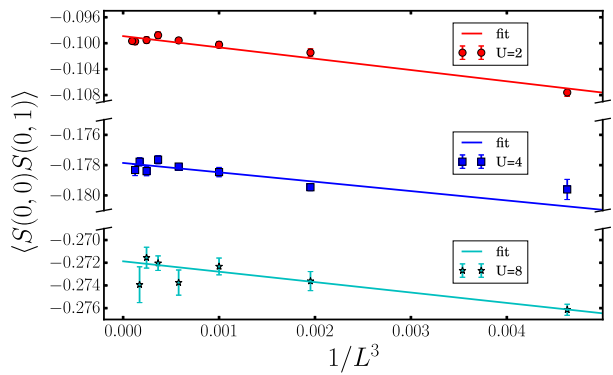}
 	\caption{Nearest neighbor (${\mathbf r} = (0,1)$) correlation function versus $1/L^3$ for $U/t = 2, 4$ and $8$.
	The largest system size included is $22 \times 22$.
 		Finite size scaling fits using Eq.~(\ref{spin_corr_scaling}) are also shown.
 	}
 	\label{corr_half_finite_U2}
 \end{figure}

 \begin{figure}[b]
 	\includegraphics[width=8.06cm]{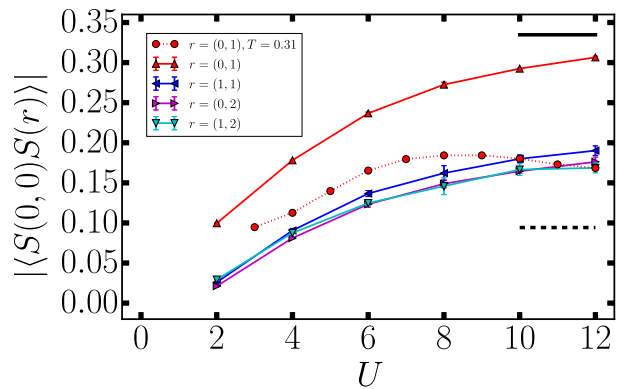}
 	\caption{The magnitude of short-range spin correlation functions at half filling. The correlation function $c({\mathbf r})$ is shown versus interaction strength, for several values of ${\mathbf r}$: $(0, 1), (1, 1), (0, 2)$, and $(1, 2)$.
	 The upper horizontal line represents the infinite-$U$ value of NN spin
 		correlation function from the Heisenberg model
		\cite{sandvik_1997}. The lower horizontal
 		line represents the infinite distance value at the infinite-$U$ limit (the square of the magnetization in the Heisenberg model \cite{sandvik_1997}). The dotted
 		red curve is the NN spin correlation function at $T = 0.31$, taken from Ref.~\cite{prl_106_035301}.}
 	\label{corr_half}
 \end{figure}

\begin{table}[h]	
	\centering
	\caption{Values of the spin correlation functions plotted in Fig.~\ref{corr_half} for $U = 2, 4, 6, 8, 10$, and $12$,
		with distance $\mathbf{r} = (0, 1), (1,1),( 0,2)$, and $(1,2)$.}
	\begin{tabular}{c|cccc}
		\hline
		\hline
		 $U/t$ & $\mathbf{r} = (0, 1)$ & $(1,1)$ & $(0,2)$ & $(1,2)$\\ \hline
		$2$& $-0.0996(7)$  &$0.0260(8)$ & $0.0211(8)$ & $-0.0294(7)$ \\ \hline
		$4$& $-0.1782(7)$  &$0.091(3)$  & $0.0811(8)$ & $-0.0875(9)$ \\ \hline
		$6$& $-0.237(2)$   &$0.137(4)$  & $0.123(4)$  & $-0.125(3)$ \\ \hline
		$8$& $-0.273(3)$   &$0.162(9)$  & $0.149(5)$  & $-0.15(1)$ \\ \hline
		$10$& $-0.293(4)$   &$0.180(5)$  & $0.165(3)$  & $-0.167(7)$ \\ \hline
		$12$& $-0.307(2)$  &$0.190(6)$  & $0.176(6)$  & $-0.169(6)$ \\ \hline
		\hline
	\end{tabular}
	\label{corr_half_table}
\end{table}

In this section, we present results for and discuss the half-filling case. 
Following the procedure in Ref.~\cite{prb_benchmark}, we take advantage of the TABC to remove errors from the finite size effect from the use of supercells. 
Lattice size up to $22 \times 22$ are studied, which is sufficient for convergence. In Fig.~\ref{corr_half_finite_U2}, we show the results of the nearest neighbor (NN), i.e., ${\mathbf r} = (0, 1)$, spin-spin correlation.
It is seen that the NN
correlation is already converged to within $1\%$ with a $14 \times 14$ lattice.
We also perform a finite-size fit of the results. Since the NN 
spin correlation function (in the infinite large U case)
is actually given by the energy in the Heisenberg model on a square lattice, we fit the short-range spin correlation function using the result from spin-wave theory \cite{finite_size_scaling_C}:
\begin{equation}
c_L = c_\infty + a/L^3
	\label{spin_corr_scaling}
\end{equation}
where $c_L$ is a spin correlation function defined in Eq.~(\ref{eqn:s(r)}) for system with size $L \times L$ and $c_\infty$ is
the  thermodynamic value. 
The quality of the fits, as shown  in Fig.~\ref{corr_half_finite_U2}, indicate that the finite-size effects in the  
NN spin 
correlation function is indeed captured well by the form in Eq.~(\ref{spin_corr_scaling}).

The spin correlation functions for all ${\mathbf r}$ between 
$(0, 1)$ and $(1, 2)$
at thermodynamic limit are plotted 
in Fig.~\ref{corr_half}. 
Spin correlations at large distances, in the context of determining the long-range order, 
 have been computed in Ref.~\cite{prb_benchmark}.
From Fig.~\ref{corr_half}, we see that the spin correlation functions always increase as $U$ is increased. The
NN 
spin correlation function approaches the Heisenberg value, $-0.334718(3)$ \cite{sandvik_1997} when $U$ approaches infinity.
Also at large $U$ the square root of the correlation function in the infinite distance limit should
approach the value of magnetization in the 2D Heisenberg limit
($0.3070(3)$ in Ref.~\cite{sandvik_1997}). We also list all the spin correlation function values in
Table~\ref{corr_half_table}.

We observe that the behavior of spin correlation functions at zero temperature is different from that at finite temperatures
where thermal fluctuation is present. For comparison, we include in Fig. \ref{corr_half} the NN spin correlation at $T = 0.31$ from Ref.~\cite{prl_106_035301}.
While the $T=0\,$K result increases monotonically with $U$, 
the $T = 0.31$  NN spin correlation reaches a maximum value at a
finite $U$ ($\sim 8$) \cite{prl_106_035301,pra_84_053611}. 
The difference  results from the competition between quantum and thermal fluctuations. 
At $T=0$\,K, with no thermal fluctuations, the increase of $U$ always drives the system towards
the Heisenberg limit where no double occupancy is allowed.
At a given non-zero temperature, however, the effective Heisenberg
anti-ferromagnetic coupling \cite{MacDonald_prb_1988}, $t^2/U$, decreases with $U$. (This implies that, in the infinite-$U$ limit, the corresponding Heisenberg
model would be at infinitely high temperature.) So we expect that the value of $U$ where the NN spin correlation function reaches its maximum will increase as 
the temperature $T$ is decreased, and become infinitely large at the limit of $T=0$\,K. 
This behavior should be easy to confirm experimentally as lower temperatures are reached 
with ultracold atoms in optical lattices.

The value of the on-site spin correlation function at half-filling can be inferred from the double occupancy results in Ref.~\cite{prb_benchmark} through
the following:
\begin{equation}
c(\mathbf{0})=\frac{3}{4}(1-2\langle n_{\uparrow}n_{\downarrow}\rangle)\,.
\label{s_0}
\end{equation}
Unlike the spin correlation function, the double occupancy always decreases with increasing $U$.
Correspondingly, the on-site spin correlation function always 
increases with interaction, both at zero
and finite temperatures.

 \begin{figure}[h]
 	\includegraphics[width=8.06cm]{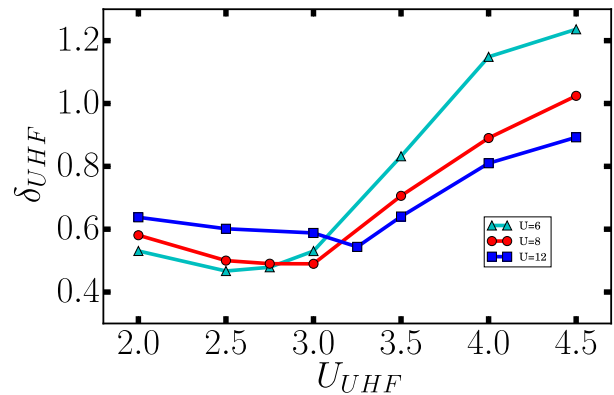}
 	\caption{Optimal value of  $U_{\rm eff}^{\rm UHF}$. The relative difference $\delta_{\text{UHF}}$ 
	is plotted versus the effective interaction used in the UHF calculation, for an 
	$8 \times 8$ system with
 		$N_\uparrow=N_\downarrow=28$  ($n = 0.875$) and $U/t = 6, 8, 12$.
 	}
 	\label{diff_U6_8_12}
 \end{figure}

 \begin{figure}[h]
 	\includegraphics[width=8.06cm]{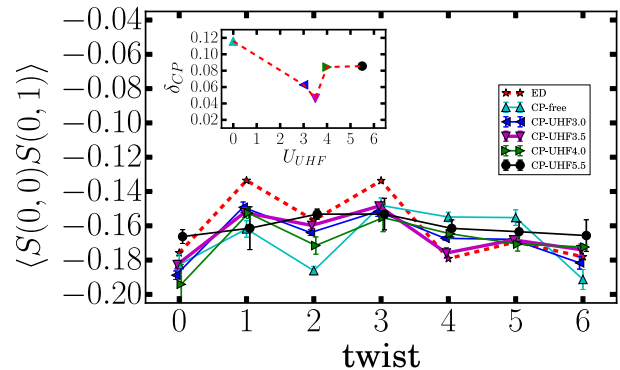}
 	\caption{Comparison of CP-AFQMC and exact diagonalization results of the NN spin correlation function for a $4 \times 4$ system with
		 $N_\uparrow=N_\downarrow=7$ and $U = 12$.
 		A series of UHF trial wave-functions, generated with different $U_{\rm eff}^{\rm UHF}$ values, 
		 are used in CP-AFQMC to evaluate the systematic errors. CP-free indicates CP-AFQMC using free-electron
		 trial wave-function. In the inset, we also plot the relative error 
		 	of the twist averaged results versus $U_{\rm eff}^{\rm UHF}$ of the UHF in trail wave-function ($\delta_{CP}$ in Eq.~(\ref{err})).
		 	The minimum error
		 	corresponds to the optimal $U_{\rm eff}^{\rm UHF} \sim 3.5$ in this case.
 	}
 	\label{4_4_7_7_C01}
 \end{figure}

\section{Results away from half-filling}
\label{away_half_filling}

\subsection{Self-consistent constraint and error quantification}

In order to control the 
sign problem 
we  a constraint
on the paths of the importance-sampled random walks in Slater determinant space,
as mentioned in Sec.~\ref{ssec:method}.
A UHF trial wave function is used, which is generated with an effective interaction, $U_{\rm eff}^{\rm UHF}$, via
a self-consistent iteration with the AFQMC calculation  \cite{prb_self-consistent}.
The optimal $U_{\rm eff}^{\rm UHF}$ is the value of
 effective $U$ with which the corresponding UHF solution yields results closest to those from AFQMC
 (using the UHF as constraining trial wave-function and always performed with the physical $U$). 
 To measure ``closeness'' we seek to minimize the following metric in the present work:
\begin{equation}
\delta_{\text{UHF}}=\frac{\sum_{\mathbf r}|c_{\text{UHF}}(\mathbf r)-c_{\text{AFQMC}}(\mathbf r)|}{\sum_{\mathbf r}|c_{\text{AFQMC}}(\mathbf r)|}\,.
\label{err_UHF_CPMC}
\end{equation}
Since we treat periodic supercells and the  UHF solution breaks translational symmetry, a subtlety arises in 
using the UHF as a trial wave-function and also in comparing the results as in Eq.~(\ref{err_UHF_CPMC})
above. One can think of this as applying a small pinning field as was the situation in Refs.~\cite{prb_self-consistent,stripe}. In the TDL one would expect the effect on the short-range correlations to be small. This is
supported by our finding below that the optimal $U_{\rm eff}^{\rm UHF}$ is slightly larger for smaller supercells
but approaches the results in  Ref.~\cite{prb_self-consistent} in large supercell sizes. It is important 
to note that the AFQMC results are insensitive to variations in the trial wave function that result
from small changes in the value of $U_{\rm eff}^{\rm UHF}$.

In Fig.~\ref{diff_U6_8_12},  we show the  $8 \times 8$ system with $U = 6, 8$ and $12$ and $n = 7/8$ as an example
for the search of the effective $U$ for the UHF trial wave-function.
The optimal $U_{\rm eff}^{\rm UHF}$ is seen to fall between $2.5$ and $3.0$ which is close to the results ($U_{\text{eff}} \sim 2.7$ for $U=8$)
in Ref.~\cite{prb_self-consistent}, where fully self-consistent calculations were performed.      
We apply this procedure to
all other fillings and $U$ values and determine the corresponding optimal $U_{\rm eff}^{\rm UHF}$  in the same way. We then
study systems with larger supercell sizes using the UHF trial wave-functions with the same 
corresponding  $U_{\rm eff}^{\rm UHF}$.
As correlation effects are diminished in systems with small $U$ and
at very low filling factors, the optimal trial wave-functions are found to become the free electron wave-functions
($U_{\rm eff}^{\rm UHF}=0$).

We also carried out an additional benchmark of CP-AFQMC, comparing results 
with exact diagonalization (ED) in the worst case scenario of the  $4 \times 4$ lattice, with  $U = 12$  and $N_\uparrow=N_\downarrow=7$. Results are shown 
in Fig.~\ref{4_4_7_7_C01}.
We plot the NN spin correlation functions for $7$ quasi-random twist angles, where the CP-AFQMC 
calculations were performed with UHF trial wave-functions generated with different $U_{\rm eff}^{\rm UHF}$.
We calculate the relative absolute error of the NN
spin correlation
with respect to the ED values as:
\begin{equation}
\delta_{\text{CP}}=\frac{\sum_{\mathbf{\Theta}}|c_{\text{AFQMC}}^{\mathbf{\Theta}}-c_{\text{ED}}^{\mathbf{\Theta}}|}{\sum_{\mathbf{\Theta}}|c_{\text{ED}}^{\mathbf{\Theta}}|}\,.
\label{err}
\end{equation}
From the inset of Fig.~\ref{4_4_7_7_C01}, we see that $U_{\rm eff}^{\rm UHF}\sim 3.5$ yields a
minimum $\delta_{\text{CP}}$, of $\sim 5\%$.
This
is consistent with the
optimal $U_{\rm eff}^{\rm UHF}$ value 
determined by the minimization of $\delta_{\text{UHF}}$ in Fig.~\ref{diff_U6_8_12}. 
As mentioned above, it is reasonable in a smaller system that $U_{\rm eff}^{\rm UHF}$ is slightly larger, 
due to a need to overcompensate for lack of dynamic correlations. 
It should be noted that the final TABC result will have a smaller error, because of error cancellations from different twist angles. (For the system shown in Fig. 4, it is 3\%.) The results in this system 
provide a kind of upper bound estimate to the CP bias.  
Lower $U$ and other doping parameters all make the method much more accurate. Larger supercell sizes 
are also expected to reduce the sensitivity of the short-range correlations. 

\subsection{Results}

\begin{figure}[b]
  	\includegraphics[width=8.06cm]{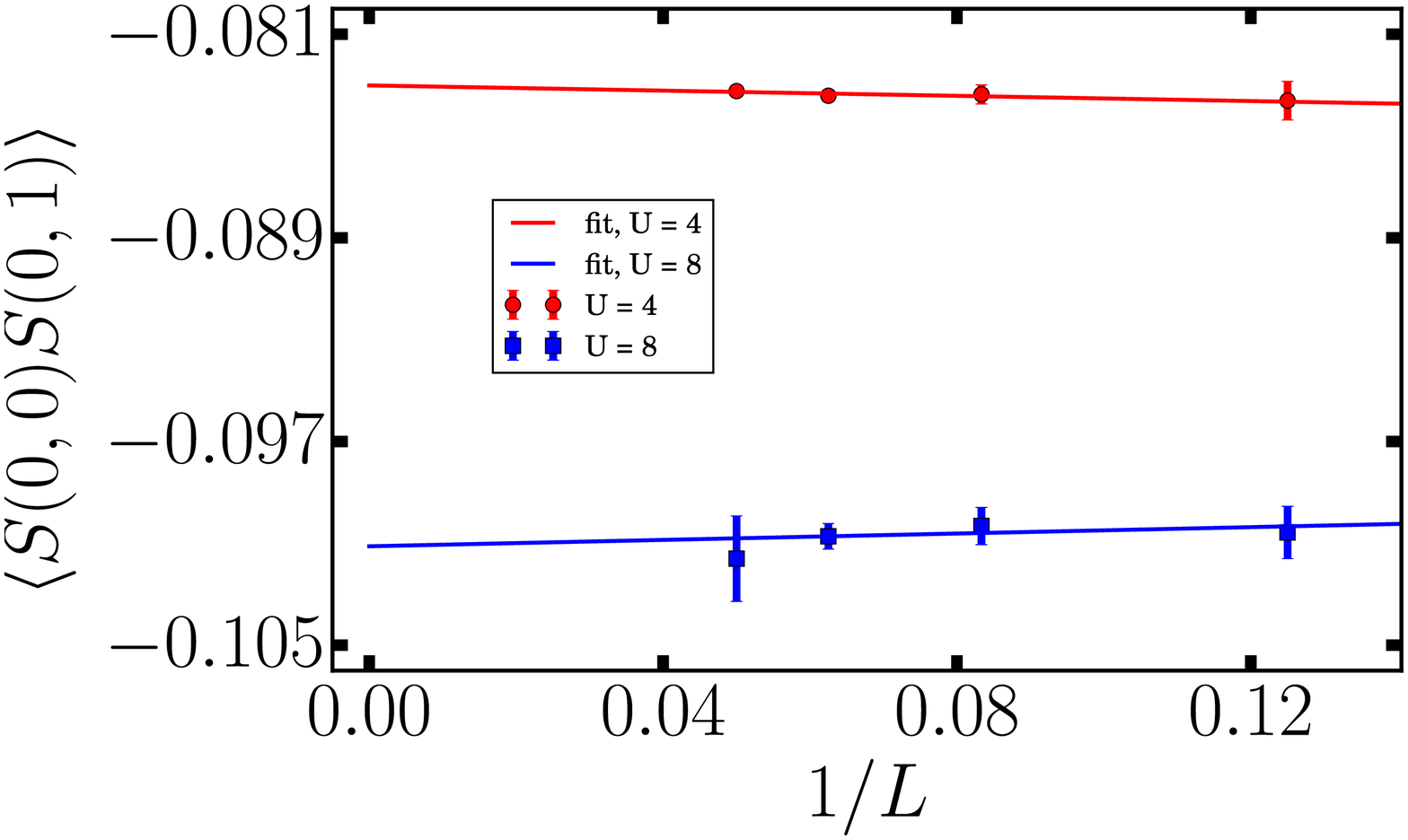}
  	\includegraphics[width=8.06cm]{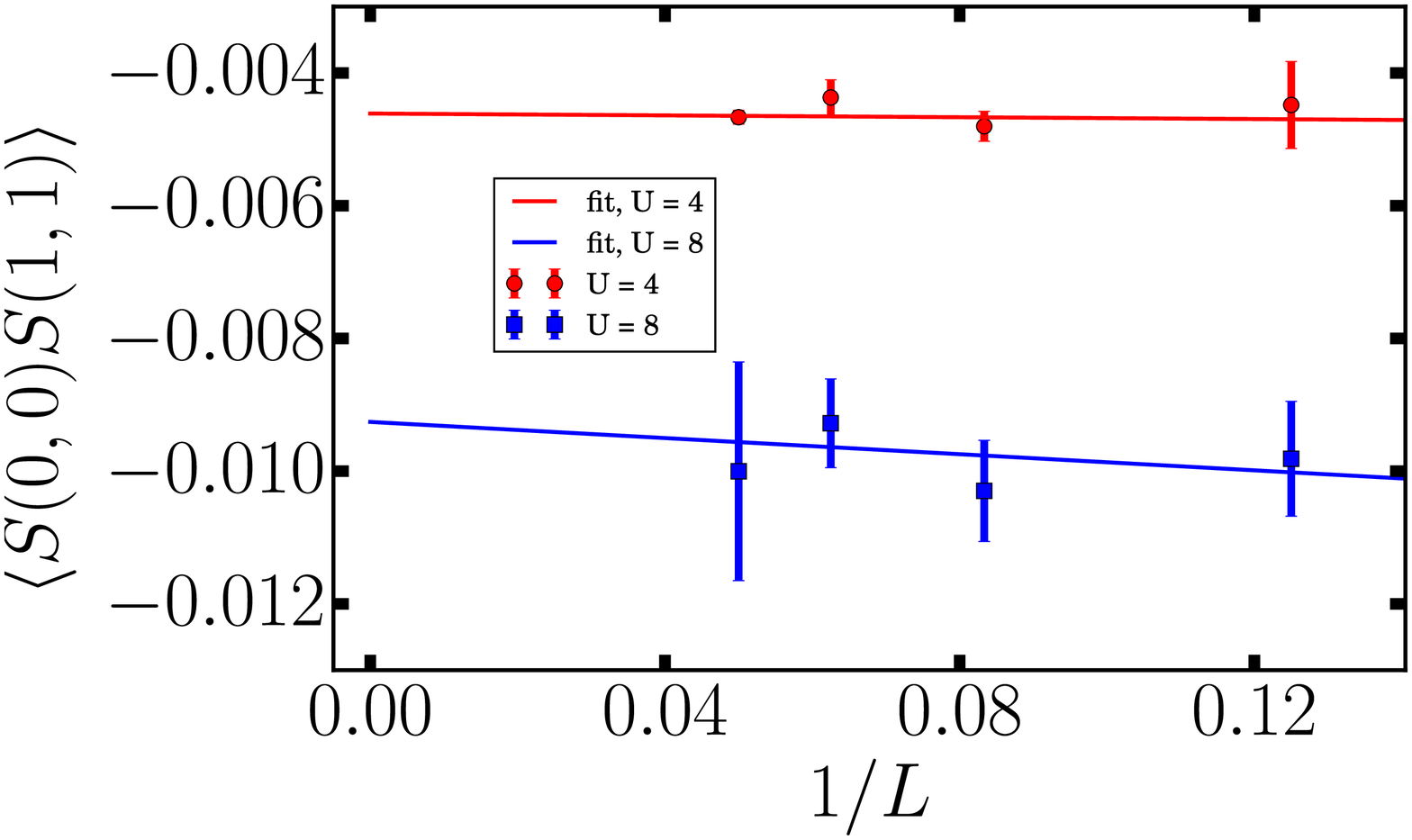}
  	\caption{Illustration of the dependence of NN (upper panel) and NNN (lower panel) spin correlation functions on supercell size. Two interaction strengths, $U = 4$ and $U = 8$, are shown in a system with $n = 0.75$. Linear fits
  		in $1/L$ are also plotted.
		}
  	\label{finite_U48_n0.5}
  \end{figure}

In all our calculations we study sufficiently large supercells to ensure that finite-size effects are negligible 
in the computed spin correlation functions.
In Fig.~\ref{finite_U48_n0.5},
we plot the NN and NNN spin correlation functions for $U = 4$ and $8$ at a filling factor of $n = 0.75$.
Linear fits of the results 
with $1/L$  are also shown. 
 It can be
seen that, in these systems, the finite size effect is smaller than the targeted statistical accuracy even at $12 \times 12$. 

  \begin{figure}[t]
  	\includegraphics[width=8.06cm]{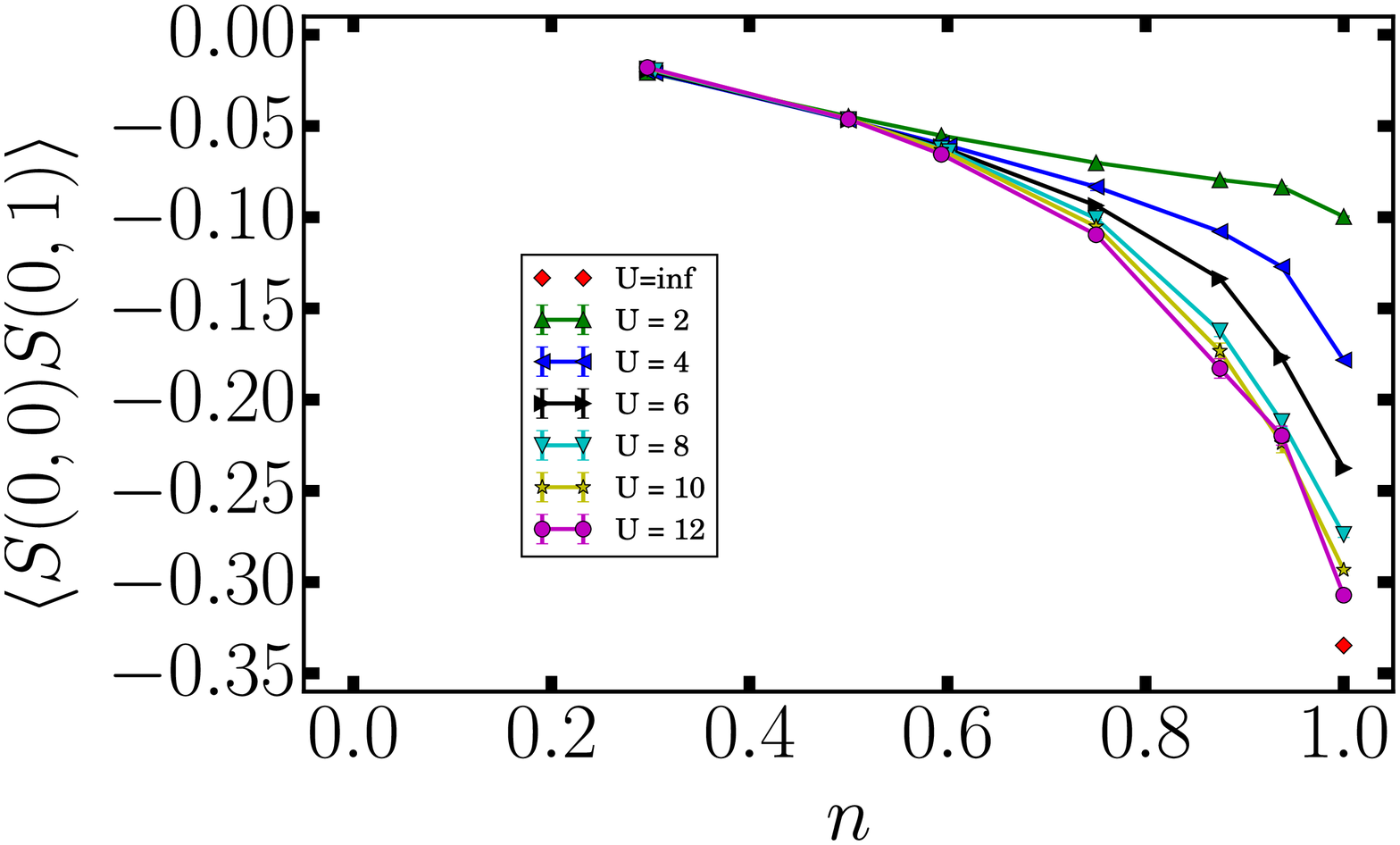}
  	\includegraphics[width=8.06cm]{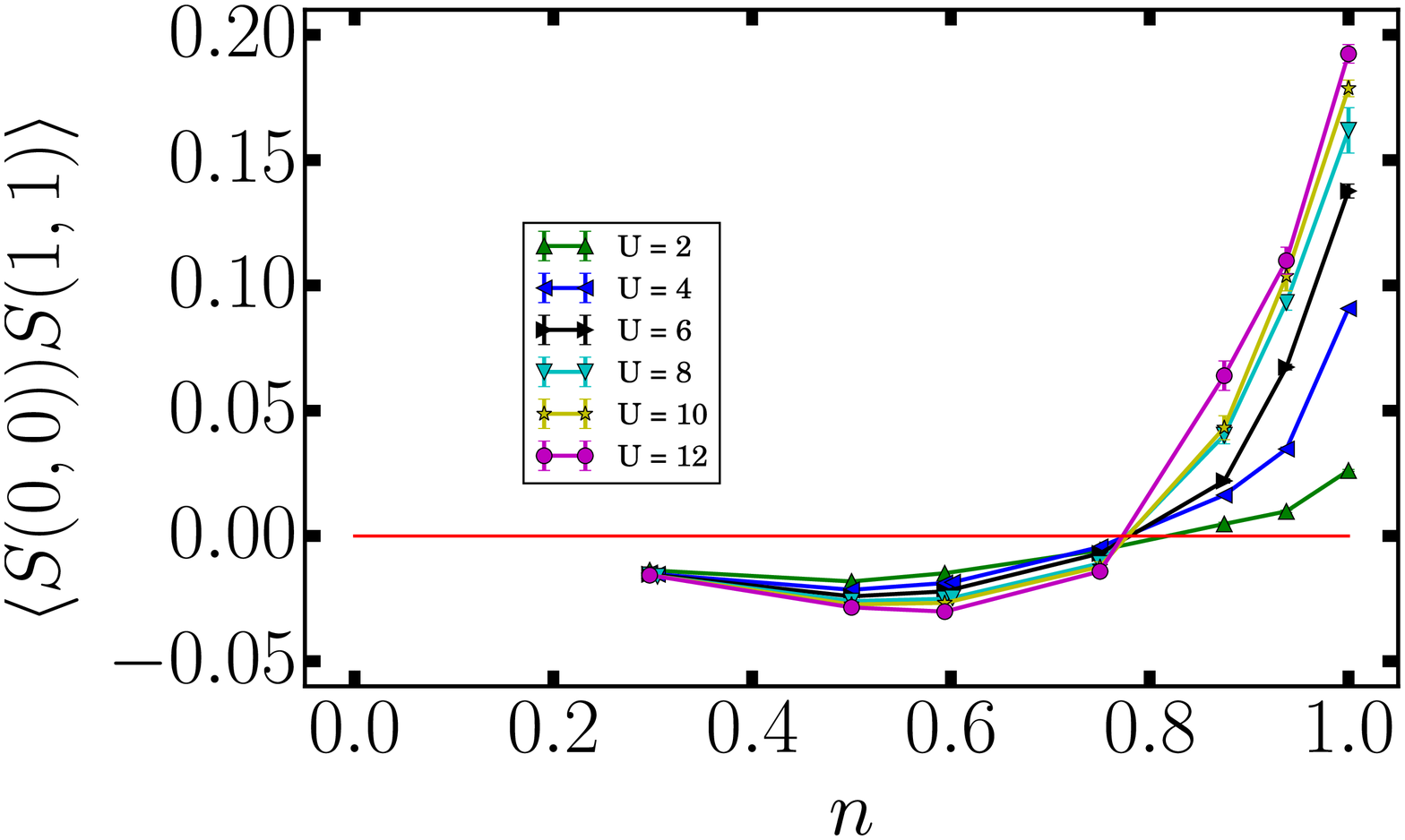}
  	\caption{NN (upper panel) and NNN (lower panel) spin correlation functions versus filling factors at different interaction strengths.
  		The half-filling results in Fig.~\ref{corr_half} are included here for completeness. The value of NN spin correlation
  		function at half-filling for $U=\infty$ is from Ref.~\cite{sandvik_1997}. 
  	}
  	\label{corr_doped}
  \end{figure}

 In Fig.~\ref{corr_doped}, we plot the NN and NNN spin correlation functions for interaction strengths ranging
 from $U = 2$ to $U = 12$.
 The negative sign of the NN spin correlation function reflects the short-range anti-ferromagnetic correlations
 in the Hubbard model. The dependence of the correlation on $U$ is mild at the dilute limit, which is
 reasonable since double occupancy is significantly reduced.  
 As the density is increased, stronger 
 dependence of the correlation on $U$ is seen from weak to moderate interactions.
 For even larger interaction strengths, the NN spin correlation functions approach the value of infinite-$U$ limit where
 no double occupancy is allowed. 
   
 For the NNN spin correlation function,  the sign is negative from the dilute limit through $n\sim 0.8$.
 This is a reflection of the exchange-correlation hole which continues from on-site to 
 NN to NNN correlations and so on. 
  Similar to 
 the NN spin correlation function,
 it only shows mild dependence on the interaction strength in the dilute limit. 
 As the filling factor is further increased,
 the NNN spin correlation
 changes sign from negative to positive, which is a precursor of the buildup of anti-ferromagnetic order in
 the system. 
 Interestingly, the crossover points in density are very close for different interaction strengths.
 The change of sign in the NNN spin correlation function, needless to say, does not necessarily 
 imply long-range
 order in the systems. To establish the existence of long-range anti-ferromagnetic orders, a more 
 systematic examination of the behavior of the tail of $c({\mathbf r})$ is necessary (see, e.g.~Ref.~\cite{prb_benchmark} for analysis at half-filling).

Previous AFQMC calculations have shown the existence of spin-density wave states at intermediate
 interaction ($U\sim 4$) for density larger than $\sim 0.85$ \cite{chang_prl_2010}. 
 For larger interactions they evolve into stripe states \cite{chang_prl_2010,stripe}.
  The determination of such phases requires the study of long-range spin correlation functions
  \cite{chang_prl_2010} or long-range density and spin-density variations in the presence of pinning 
  fields  \cite{stripe}, as well as careful removal of the influence of finite size  and 
  supercell shape \cite{chia-chen_EOS,chang_prl_2010,stripe}.
   The effect of these collective modes in the ground state is less direct in the short-range correlations 
 computed here, although clearly they can lead to quantitative modifications (reductions) in the magnitude of NN
 and NNN spin correlations. The slopes of the curves in Fig.~\ref{corr_doped} show large increases in magnitude as we approach half-filling, which is consistent with and could be a manifestation of these states.

\section{Summary}
\label{summary}
In summary, we have calculated the short-range spin correlation functions in the two-dimensional Hubbard model at zero temperature with AFQMC.
At half-filling, the results are numerically exact. The absolute values of spin correlation functions are found to always increase
with $U$, which is different from the finite temperature behavior. 
Away from half-filling, we eliminate the sign problem with a 
 self-consistent constraint.
The systematic errors from the constraint are examined and quantified. 
We observe a change of sign from negative to positive in the next nearest neighbor spin correlation function
as a function of doping, 
with a crossing point
slightly below $0.8$ which shows very weak dependence on the
 interaction strength. 
The results in this paper can serve as a valuable 
benchmark as optical lattice experiments with ultra-cold atoms reach lower temperatures.
The detailed and quantitative results on spin correlation functions provide useful information
in the search for theoretical understanding and experimental realization of exotic phases 
of magnetic and accompanying or competing charge and possibly superconducting orders.  
 
\begin{acknowledgments}
We acknowledge  support from NSF (DMR-1409510). 
MQ and SZ were also supported by the Simons Foundation.
The calculations were carried out at
the Extreme Science and Engineering Discovery Environment
(XSEDE), which is supported by National Science Foundation grant number ACI-1053575,
and the computational facilities at the College of William and Mary. We
gratefully acknowledge a Director's discretionary allocation at OLCF.
\end{acknowledgments}


\begin{thebibliography}{1}
	
\bibitem{Hubbard_origional}
J. Hubbard, Proceedings of the Royal Society of London
A: Mathematical, Physical and Engineering Sciences {\bf 276}, 238 (1963).

\bibitem{Imada_rmp_1998}
Masatoshi~Imada, Atsushi~Fujimori, and Yoshinori~Tokura,
Rev. Mod. Phys. {\bf 70}, 1039(1998).

\bibitem{cdw_sdw}
Jie~Xu, Chia-Chen~Chang, Eric~J.~Walter, Shiwei~Zhang,
J. Phys.: Condens. Matter {\bf 23}, 505601 (2011), 
Robert~Peters and Norio~Kawakami, Phys. Rev. B {\bf 89}, 155134(2014) 

\bibitem{ph-sym}
J.~E.~Hirsch, Phys. Rev. B {\bf 31}, 4403 (1985).

\bibitem{htc}
P. W. Anderson, Science {\bf 235}, 1196 (1987),
Elbio~Dagotto, Rev. Mod. Phys. {\bf 66}, 763 (1994).

\bibitem{lieb_wu}
E. H. Lieb and F. Y. Wu, Phys. Rev. Lett. {\bf 20}, 1445 (1968).

\bibitem{paper_simons}
J. P. F. LeBlanc, Andrey E. Antipov, Federico Becca, Ireneusz W. Bulik, Garnet Kin-Lic Chan, Chia-Min Chung, Youjin Deng, Michel Ferrero, Thomas M. Henderson, Carlos A. Jimenez-Hoyos, E. Kozik, Xuan-Wen Liu, Andrew J. Millis, N. V. Prokofev, Mingpu Qin, Gustavo E. Scuseria, Hao Shi, B. V. Svistunov, Luca F. Tocchio, I. S. Tupitsyn, Steven R. White, Shiwei Zhang, Bo-Xiao Zheng, Zhenyue Zhu, and Emanuel Gull, Phys. Rev. X {\bf 5}, 041041(2015).

\bibitem{rev_zoller}
D. Jakscha, P. Zoller, Ann. Phys {\bf 315}, 52 (2005).

\bibitem{rev_bloch}
Immanuel Bloch, Jean Dalibard, and Wilhelm Zwerger, Rev. Mod. Phys. {\bf 80}, 885 (2008).

\bibitem{nat_519_2015}
Russell A. Hart, Pedro M. Duarte, Tsung-Lin Yang, Xinxing Liu, Thereza Paiva, Ehsan Khatami, Richard T. Scalettar, Nandini Trivedi,	David A. Huse, and Randall G. Hulet, Nature {\bf 519}, 211 (2015).

\bibitem{prl_116_175301}
Eugenio Cocchi, Luke A. Miller, Jan H. Drewes, Marco Koschorreck, Daniel Pertot, Ferdinand Brennecke, and Michael Kohl, Phys. Rev. Lett. {\bf 116}, 175301 (2016).

\bibitem{prl_116_235301}
Lawrence W. Cheuk, Matthew A. Nichols, Katherine R. Lawrence, Melih Okan, Hao Zhang, and Martin W. Zwierlein, Phys. Rev. Lett. {\bf 116}, 235301 (2016).

\bibitem{sci_1253}
Maxwell F. Parsons, Anton Mazurenko, Christie S. Chiu, Geoffrey Ji, Daniel Greif, Markus Greiner, Science {\bf 353}, 1253 (2016).

\bibitem{sci_1257}
Martin Boll, Timon A. Hilker, Guillaume Salomon, Ahmed Omran, Jacopo Nespolo, Lode Pollet, Immanuel Bloch, Christian Gross, Science {\bf 353}, 1257 (2016).

\bibitem{sci_1260}
Lawrence W. Cheuk, Matthew A. Nichols, Katherine R. Lawrence, Melih Okan, Hao Zhang, Ehsan Khatami, Nandini Trivedi, Thereza Paiva, Marcos Rigol, Martin W. Zwierlein, Science {\bf 353}, 1260 (2016).

\bibitem{stripe}
Bo-Xiao Zheng, Chia-Min Chung, Philippe Corboz, Georg Ehlers, Ming-Pu Qin, Reinhard M. Noack, Hao Shi, Steven R. White, Shiwei Zhang, Garnet Kin-Lic Chan,  arXiv:1701.00054 (2017).



\bibitem{prl_104_066406}
Thereza Paiva, Richard Scalettar, Mohit Randeria, and Nandini Trivedi, Phys. Rev. Lett. {\bf 104}, 066406 (2010).

\bibitem{prl_106_035301}
Simone Chiesa, Christopher N. Varney, Marcos Rigol, and Richard T. Scalettar, Phys. Rev. Lett. {\bf 106}, 035301 (2011).

\bibitem{prb_88_155108}
J. P. F. LeBlanc and Emanuel Gull, Phys. Rev. B {\bf 88}, 155108  (2013).

\bibitem{pra_84_053611}
Ehsan Khatami and Marcos Rigol, Phys. Rev. A {\bf 84}, 053611 (2011).

\bibitem{prl_109_205301}
Baoming Tang, Thereza Paiva, Ehsan Khatami, and Marcos Rigol, Phys. Rev. Lett. {\bf 109}, 205301  (2012).

\bibitem{Mazurenko_2016}
Anton Mazurenko, Christie S. Chiu, Geoffrey Ji, Maxwell F. Parsons, Márton Kanász-Nagy, Richard Schmidt, Fabian Grusdt, Eugene Demler, Daniel Greif, Markus Greiner,
arXiv:1612.08436 (2016).

\bibitem{sign_1}
E.~Y.~Loh Jr., J.~E.~Gubernatis, R.~T.~Scalettar, S.~R.~White, D.~J.~Scalapino, and R.~L.~Sugar, Phys. Rev. B {\bf 41}, 9301 (1990).

\bibitem{sign_2}
K. E. Schmidt and M. H. Kalos, in Applications of the Monte Carlo Method in Statistical Physics, edited by K. Binder (Springer-Verlag, Heidelberg, 1984).

\bibitem{zhang_prb_1997}
S.~Zhang, J.~Carlson, and J.~E.~Gubernatis, Phys. Rev.
B {\bf 55}, 7464 (1997).

\bibitem{chia-chen_EOS}
Chia-Chen Chang and Shiwei Zhang, Phys. Rev. B {\bf 78}, 165101 (2008).

\bibitem{CPMC_sym_1}
Hao Shi and Shiwei Zhang, Phys. Rev. B {\bf 88}, 125132 (2013).

\bibitem{prb_self-consistent}
Mingpu Qin, Hao Shi, Shiwei Zhang, Phys. Rev. B {\bf 94}, 235119 (2016).

\bibitem{TBC}
C.~Lin, F.~H.~Zong and D.~M.~Ceperley, Phys. Rev. E {\bf 64}, 016702 (2001).

\bibitem{prb_benchmark}
Mingpu Qin, Hao Shi, Shiwei Zhang, Phys. Rev. B {\bf 94}, 085103 (2016).

\bibitem{lecture-notes}
S. Zhang, Auxiliary-Field Quantum Monte Carlo for Correlated Electron Systems, Vol. 3 of Emergent Phenomena in Correlated Matter: Modeling and Simulation, Ed. E. Pavarini, E. Koch, and U. Schollw¨ock (Verlag des Forschungszentrum J¨ulich, 2013).

\bibitem{AFQMC}
R.~Blankenbecler, D.~J.~Scalapino and R.~L.~Sugar, Phys. Rev. D {\bf 24}, 2278 (1981).

\bibitem{koonin}
G. Sugiyama and S. E. Koonin, Ann. Phys. (N.Y.) {\bf 168}, 1 (1986).

\bibitem{TREMBLAY_1992}
Liang Chen,  A.-M. S. Tremblay, Int. J. Mod. Phys. B {\bf 06}, 547 (1992).

\bibitem{hirsch_prb_1983}
J.~E.~Hirsch,
Phys. Rev. B {\bf 28}, 4059 (1983).

\bibitem{CPMC_sym_2}
Hao Shi, Carlos A.~Jimenez-Hoyos, R.~Rodriguez-Guzman, Gustavo E.~Scuseria, and Shiwei Zhang, Phys. Rev. B {\bf 89}, 125129 (2014).

\bibitem{Hao-inf-var}
Hao Shi and Shiwei Zhang, Phys. Rev. E {\bf 93}, 033303 (2016).


\bibitem{hao_pra_2015}
H. Shi, S. Chiesa, and S. Zhang, Phys. Rev. A {\bf 92}, 033603 (2015).


\bibitem{zhang_prl_2003}
S. Zhang and H. Krakauer, Phys. Rev. Lett. 90, 136401 (2003).




\bibitem{Huy_CPC_2014}
Huy Nguyen, Hao Shi, Jie Xu, Shiwei Zhang, Comput. Phys. Commun {\bf 185}, 3344 (2014).

\bibitem{Wirawan-PRE}
Wirawan Purwanto and Shiwei Zhang, Phys. Rev. E {\bf 70}, 056702 (2004).



\bibitem{finite_size_scaling_C}
Herbert Neuberger and Timothy Ziman, Phys. Rev. B {\bf 39}, 2608 (1989), Daniel S.~Fisher, Phys. Rev. B {\bf 39}, 11783 (1989).

\bibitem{sandvik_1997}
Anders~W.~Sandvik, Phys. Rev. B {\bf 56}, 11678 (1997).

\bibitem{MacDonald_prb_1988}
A.~H.~MacDonald, S.~M.~Girvin and D.~Yoshioka, Phys. Rev. B {\bf 37}, 9753 (1988).

\bibitem{chang_prl_2010}
C.-C. Chang and S. Zhang, Phys. Rev. Lett. 104, 116402 (2010).

	
\end{thebibliography}
\end{document}